# Three-dimensional structure determination from a single view

Kevin S. Raines[1], Sara Salha[1], Richard L. Sandberg[2], Huaidong Jiang[1], Jose A. Rodríguez[3], Benjamin P. Fahimian[1], Henry C. Kapteyn[2], Jincheng Du[4] and Jianwei Miao[1]

*[1]Department of Physics and Astronomy and California NanoSystems Institute, University of California, Los Angeles, CA 90095, USA. [2]Department of Physics and JILA, University of Colorado and NIST, Boulder, CO 80309, USA. [3]Molecular Biology Institute, University of California at Los Angeles, Los Angeles, CA 90095, USA. [4]Deparment of Materials Science & Engineering and Center for Advanced Scientific Computing and Modeling, University of North Texas, Denton, TX 76203, USA*

**The ability to determine the structure of matter in three dimensions has profoundly advanced our understanding of nature. Traditionally, the most widely used schemes for 3D structure determination of an object are implemented by acquiring multiple measurements over various sample orientations, as in the case of crystallography and tomography[1,2], or by scanning a series of thin sections through the sample, as in confocal microscopy[3]. Here we present a 3D imaging modality, termed ankylography (derived from the Greek words *ankylos* meaning 'curved' and *graphein* meaning 'writing'), which enables complete 3D structure determination from a single exposure using a monochromatic incident beam. We demonstrate that when the diffraction pattern of a finite object is sampled at a sufficiently fine scale on the Ewald sphere, the 3D structure of the object is determined by the 2D spherical pattern. We confirm the theoretical analysis by**

**performing 3D numerical reconstructions of a sodium silicate glass structure at 2 Å resolution and a single poliovirus at 2 – 3 nm resolution from 2D spherical diffraction patterns alone. Using diffraction data from a soft X-ray laser, we demonstrate that ankylography is experimentally feasible by obtaining a 3D image of a test object from a single 2D diffraction pattern. This approach of obtaining complete 3D structure information from a single view is anticipated to find broad applications in the physical and life sciences. As X-ray free electron lasers (X-FEL) and other coherent X-ray sources are under rapid development worldwide, ankylography potentially opens a door to determining the 3D structure of a biological specimen in a single pulse and allowing for time-resolved 3D structure determination of disordered materials.**

Lensless imaging techniques generally begin with the coherent diffraction pattern of a non-crystalline sample, which is measured and then directly phased to obtain an image. The initial insight that the continuous diffraction pattern of a non-crystalline object might be invertible was suggested by Sayre in 1980 (ref. 4). In 1999, coherent diffraction microscopy was experimentally demonstrated by Miao *et al*.[5]. Since that initial demonstration, lensless imaging has advanced considerably, and has been applied to a wide range of samples including nanoparticles, nanocrystals, biomaterials, cells, cellular organelles, viruses and carbon nanotubes by synchrotron radiation[6-19], electrons[20,21], optical lasers[21,22], high harmonic and short wavelength laser sources[23,24], and free electron lasers[25-27]. However, in order to generate 3D images by coherent diffraction microscopy, multiple diffraction patterns are required at different sample orientations[7,10-13,16,19]. The precise mechanical tilting necessary to obtain these patterns prevents 3D imaging by a single X-FEL pulse or time-resolved 3D structure determination of disordered materials. Here we develop a novel imaging modality, ankylography that overcomes these limitations.

Fig. 1 shows a schematic layout of the experimental set-up for ankylography. A coherent beam of wavelength λ illuminates a finite object. The scattered waves form a diffraction pattern on the Ewald sphere, which can be expressed in the spherical polar coordinate system as (supplementary discussion)

$$|F(\theta,\varphi)| = \left| \int_V \rho(\vec{r}) e^{-\frac{2\pi i}{\lambda}[x\sin\theta\cos\varphi + y\sin\theta\sin\varphi + z(\cos\theta - 1)]} d^3\vec{r} \right| \quad (1)$$

where $|F(\theta,\varphi)|$ is proportional to the magnitude of the scattered waves, $\theta$ and $\varphi$ the spherical polar coordinates, $\rho(\vec{r})$ the 3D structure of the object with $\vec{r} = (x, y, z)$, and $V$ the spatial extent of the object. When the 2D spherical diffraction pattern is sampled at a scale sufficiently finer than the Nyquist interval such that the oversampled degree ($O_d$) is larger than 1, the oversampled diffraction pattern in principle determines the 3D structure of the object (supplementary discussion), which can be retrieved by iterative algorithms. Here $O_d$ is defined as the ratio of the number of measured intensity points from Eq. (1) to the number of unknown variables of the 3D object array[28]. Although Eq. (1) assumes plane wave illumination, ankylography can be extended to divergent or convergent wave illumination through deconvolution with the illumination function. The 3D spatial resolution of ankylography is determined by (supplementary Fig. S1)

$$d_x = d_y = \frac{\lambda}{\sin(2\theta_{max})} \qquad d_z = \frac{\lambda}{2\sin^2\theta_{max}} \quad (2)$$

where $d_x$, $d_y$ and $d_z$ represent the resolution along the X, Y and Z-axes, respectively. The Z-axis is along the beam propagation direction and $2\theta_{max}$ is the diffraction angle shown in Fig. 1.

To confirm our theoretical analysis, we conducted numerical simulations on 3D structure determination of sodium silicate glasses ($25Na_2O$-$75SiO_2$). Sodium silicate glasses are archetypal for a wide variety of multicomponent silicate glasses that find applications from glassware and window panes, to optical fibers, photonic devices and bioactive glasses. Unlike crystalline materials, glasses lack long-range periodicity, which makes experimental determination of the glass structure difficult by traditional diffraction methods. Thus the inherent complexity of glass structures combined with the lack of experimental methods to probe them (especially for medium range structures) makes understanding and visualizing the atomic and nanostructure of glasses a great challenge to the scientific community. By combining classical molecular dynamics (MD) simulations with subsequent geometry optimization using *ab initio* density functional theory calculations, we obtained a 3D sodium silicate glass structure of 204 atoms, with a size of $14 \times 14 \times 14$ Å$^3$ (supplementary methods). A beam of coherent X-rays with a wavelength of 2 Å was incident on the nanostructure. The total coherent flux illuminating the glass structure was about $10^{13}$ photons. An oversampled 2D diffraction pattern was calculated on the Ewald sphere, with $2\theta_{max} = 90°$. Poisson noise was added to the spherical diffraction pattern, shown in Fig. 1a.

To reconstruct the 3D glass structure, we embedded the 2D spherical diffraction pattern into a 3D array of $64 \times 64 \times 64$ voxels, corresponding to $O_d = 2.7$. Since the electron density of the glass structure is real, we also embedded the centro-symmetrical diffraction pattern into the same array. All other data points in the 3D array were set as unknowns. The 3D image reconstruction was carried out using an iterative phase retrieval algorithm (methods summary and supplementary discussion). Fig. 1b shows the 3D atomic positions of the reconstructed sodium silicate glass structure with a resolution of 2 Å in all three dimensions (see supplementary Fig. S2 for resolution quantification). In Fig. 1b, the red, purple and yellow spheres represent the positions of O, Na and Si atoms, respectively. Compared to the MD simulated structure, the 3D atomic positions of the sodium silicate glass structure are resolved. Fig. 2 shows two 2-Å-thick slices of the MD simulations (Figs. 2a and c) and the reconstructed structures (Figs. 2b and d) along the XY and XZ planes, respectively. The electron density distribution of the reconstructed structure is in good agreement with that of the MD simulation. The slight difference of the electron density between the two structures is likely due to the Poisson noise added to the diffraction pattern.

To explore the biological application of ankylography, we performed a numerical experiment on 3D image reconstruction of an individual poliovirus from a single X-FEL pulse. Poliovirus contains a single-stranded RNA genome within an icosahedral capsid that is delivered into a host cell via interaction of virus coat proteins with a poliovirus receptor[29]. During this process, the virus undergoes an irreversible conformational change that results in an increased affinity for the poliovirus receptor, producing what is known as the 135S particle[29]. In the numerical experiment, we simulated an X-FEL pulse with $\lambda$ = 1.77 nm and $10^{13}$ photons per pulse that was focused to a 100 nm spot and illuminated a single 135S particle. The 3D model structure of the poliovirus has been obtained by cryo-electron microscopy through averaging 8224 particles[29]. A diffraction pattern on the Ewald sphere was obtained with a diffraction angle of 62.6°, which corresponds to a spatial resolution of 2.0 nm in the XY plane and 3.3 nm along the Z-axis. Poisson noise was added to the diffraction intensities. The 3D ankylographic reconstruction was carried out without using *a priori* information of the virus structure (methods summary and supplementary discussion).

Figs. 3a and b show an iso-surface rendering and a 1.65-nm-thick central slice of the 3D virus structure reconstructed from the 2D noisy diffraction pattern alone, which are in an excellent agreement with the model (supplementary Fig. S3). The reconstructed 3D capsid structure clearly resolves the star-shaped mesa that forms the fivefold symmetry (red dots in Fig. 3a) and the propeller tips (blue dots in Fig. 3a) that form the threefold symmetry. In the vicinity of the fivefold mesa and the propeller structure is a canyon (indicated with arrows in Figs. 3a and b), which is the expected binding site of the poliovirus receptor. Compared to cryo-electron microscopy, which enables 3D imaging of large protein complexes such as viruses by averaging over thousands of particles, ankylography can in principle be used to determine the 3D *pleomorphic* structure of a single biological complex at high resolution.

To verify ankylography with experimental data, we used a 2D diffraction pattern taken with a soft X-ray laser of $\lambda = 47$ nm and $\lambda/\Delta\lambda \approx 10^4$ (ref. 24). The test pattern was fabricated by etching through a substrate consisting of a silicon nitride membrane of ~100 nm thick. Except for the patterned area, the substrate was completely opaque to the 47 nm light. A slant of the test sample relative to the incident beam served to increase the Z-axis width of the sample, providing a 3D depth (Fig. 4b). The CCD detector (Andor, 2048×2048 pixel EUV detector array, 13.5 µm × 13.5 µm pixel size) was positioned at a distance of 14.5 mm from the sample. A 2D diffraction pattern was measured by the detector with intensities extending to the edges of the CCD camera. To enhance the signal to noise ratio, we integrated the diffraction intensities by binning 3×3 pixels into 1 pixel and then interpolated the planar diffraction pattern onto the Ewald sphere (supplementary discussion, Fig. 4a). The spherical diffraction pattern was embedded into a 3D array of 420×420×240 voxels with $O_d = 2.6$.

The phase retrieval was carried out using similar methods to those used for the sodium silicate glass structure (methods summary and supplementary discussion). Figs. 4b and c show iso-surface renderings of the 3D reconstruction in the XY plane (front view) and the XZ plane (side view), respectively, in which the front view is in good agreement with the SEM image (Fig. 4e). The spatial resolution of the reconstructed image was estimated to be 80 nm along the X and Y-axes and 140 nm along the Z-axis. Based on the 3D reconstruction, we measured the tilt angle of the sample relative to the beam to be 5.1° (Fig. 4b). Fig. 4d shows a lineout along the Z-axis of the reconstructed image. The width of the tilted sample's projection onto the Z-axis was measured to be 405 nm in the reconstruction, which is in good agreement with the expected value of 389 nm computed from the sample geometry and the title angle. Moreover, we identified two structure defects in the SEM image of the test sample (arrows in Fig. 4e). The structure defects are spatially resolved in the 3D reconstructed image (arrows in Fig. 4b), which further confirms the robustness of our reconstruction.

In conclusion, we have discovered that the 3D structure information of a finite object is encoded into a 2D diffraction pattern on the Ewald sphere. When the diffraction pattern is sufficiently oversampled, the 3D structure can be directly retrieved from the 2D spherical diffraction pattern alone. Both our numerical simulations and experimental results have verified the feasibility of this novel 3D imaging technique. Compared to conventional 3D coherent diffraction microscopy[7,10-13,16,19], ankylography requires a comparable amount of incident flux for achieving a desired resolution, but redistributes all the intensity points more finely on the Ewald sphere, which eliminates the necessity of sample tilting. On the other hand, due to the oversampling requirement of the single 2D diffraction pattern, ankylography demands area detectors with a larger number of pixels to compute 3D reconstructions of sizable specimens at high resolutions. Furthermore, to reduce the interpolation errors (supplementary discussion), spherical area detectors are more desirable in ankylography. As detector technologies have been advancing significantly, these types of detectors can be developed with currently existing technologies. Finally, our numerical simulations have indicated that, by using two or more 2D spherical diffraction patterns (*i.e.* two or more sample tilts), the 3D image reconstruction process (methods summary) can be significantly sped up. Thus ankylography can also be used to considerably reduce the number of tilts required in 3D coherent diffraction microscopy.

Looking forward, by properly choosing the wavelength of the monochromatic incident beam, ankylography can in principle be implemented using synchrotron radiation, high harmonic and short wavelength laser sources, optical lasers, electrons, and X-FELs. Thus we expect the applications of ankylography to be broad across several disciplines. In biology, as optical diffraction microscopy has already been applied to imaging a biological sample[22], ankylography may open a door for 4D optical imaging of biological specimens since sample tilting, scanning or sectioning can be avoided. With X-FELs, ankylography will enable 3D structure determination of a biological specimen from a single pulse before the specimen is destroyed[25,30]. In materials science, ankylography can in principle be applied to investigating time-resolved 3D structure of disordered materials using X-FELs.

**METHODS SUMMARY**

3D image reconstruction in ankylography proceeds by means of an iterative algorithm. A random phase set defines an initial input and iterates back and forth between Fourier and real space where physical constraints were applied. In Fourier space, the magnitudes of the Fourier transform on the Ewald sphere were set to the measured values. The data points not on the Ewald sphere were initially set as unknowns and updated with each iteration of the reconstruction. In real space, the following constraints were enforced: the electron density must be non-negative, continuous, and bounded within the spatial extent of the sample (*i.e.* a support). Data points of the image estimate that did not satisfy the positivity or boundedness constraints were modified by the operations described further in the supplementary information. In order to enforce continuity of the image within the support, a Gaussian filter was periodically applied to the image density throughout the reconstruction. Outside the support, where the density should be uniform, in addition to zero-valued, convolution with a box kernel was implemented as described in the supplementary information. An error metric, defined as the difference between the computed data points on the Ewald sphere and the measured ones, was used to monitor the convergence of the algorithm.

In order to capitalize on the fast convergence of the 3D reconstructions, where knowledge of a tight support is not necessary, we begin with a low-resolution reconstruction in order to recover the missing magnitudes in Fourier space. We subsequently include these recovered amplitudes in the higher resolution reconstructions. We call this type of constraint *amplitude extension*. This constraint allows the object envelope to be easily computed without *a priory* information, an important step in the reconstruction process. A detailed description of how these constraints are implemented can be found in the supplementary information.

**References**

1. Giacovazzo, C. *et al*. *Fundamentals of Crystallography*. Oxford University Press, USA, 2nd edition (2002).

2. Kak, A.C. & Slaney, M. *Principles of Computerized Tomographic Imaging*. (SIAM, Philadelphia, 2001).

3. *Handbook of Biological Confocal Microscopy*. Edited by J. B. Pawley, 3rd Ed. (Springer, 2006).

4. Sayre, D. in *Imaging Processes and Coherence in Physics* vol. **112**, (eds Schlenker, M. *et al*.) 229-235 (Lecture Notes in Physics, Springer, 1980).

5. Miao, J., Charalambous P., Kirz, J. & Sayre, D. Extending the methodology of X-ray crystallography to allow imaging of micrometre-sized non-crystalline specimens. *Nature* **400**, 342 (1999).

6. Robinson, I. K., Vartanyants, I. A. ,Williams, G. J., Pfeifer, M. A. & Pitney, J. A. Reconstruction of the Shapes of Gold Nanocrystals Using Coherent X-Ray Diffraction. *Phys. Rev. Lett.* **87**, 195505 (2001).

7. Miao, J. *et al.* High Resolution 3D X-Ray Diffraction Microscopy. *Phys. Rev. Lett.* **89**, 088303 (2002).

8. Miao, J., Hodgson, K. O., Ishikawa, T., Larabell, C. A., LeGros, M. A. & Nishino, Y. Imaging whole Escherichia coli bacteria by using single-particle x-ray diffraction. *Proc. Natl. Acad. Sci. USA* **100**, 110 (2003).

9. Nugent, K. A., Peele, A. G., Chapman, H. N. & Mancuso, A. P. Unique Phase Recovery for Nonperiodic Objects. *Phys. Rev. Lett.* **91**, 203902 (2003).

10. Shapiro, D. *et al.* Biological imaging by soft x-ray diffraction microscopy, *Proc. Natl. Acad. Sci. USA* **102**, 15343-15346 (2005).

11. Pfeifer, M. A., Williams, G. J., Vartanyants, I. A., Harder, R. & Robinson, I. K. Three-dimensional mapping of a deformation field inside a nanocrystal. *Nature* **442**, 63 (2006).

12. Miao, J. *et al.* Three-dimensional GaN-$Ga_2O_3$ core shell structure revealed by x-ray diffraction microscopy. *Phys. Rev. Lett.* **97**, 215503 (2006).

13. Chapman, H. N. *et al.* High resolution *ab initio* three-dimensional x-ray diffraction microscopy. *J. Opt. Soc. Am. A.* **23**, 1179-200 (2006).

14. Williams, G. J. *et al.* Fresnel Coherent Diffractive Imaging. *Phys. Rev. Lett.* **97**, 025506 (2006).

15. Abbey, B. *et al.* Keyhole coherent diffractive imaging. *Nature Phys.* **4**, 394 (2008).

16. Jiang H. *et al.* Nanoscale Imaging of Mineral Crystals inside Biological Composite Materials Using X-Ray Diffraction Microscopy. *Phys. Rev. Lett.* **100**, 038103 (2008).

17. Thibault, P. *et al.* High-Resolution Scanning X-ray Diffraction Microscopy. *Science* **321**, 379-382 (2008).

18. Song, C. *et al.* Quantitative Imaging of Single, Unstained Viruses with Coherent X-rays. *Phys. Rev. Lett*. **101**, 158101 (2008).

19. Nishino, Y., Takahashi Y., Imamoto N., Ishikawa, T. & Maeshima, K. Three-Dimensional Visualization of a Human Chromosome Using Coherent X-Ray Diffraction. *Phys. Rev. Lett.* **102**, 018101 (2009).

20. Zuo, J. M., Vartanyants, I., Gao M., Zhang R. & Nagahara L. A. Atomic Resolution Imaging of a Carbon Nanotube from Diffraction Intensities. *Science* **300**, 1419-1421 (2003).

21. Spence J. C. H., Weierstall U. & Howells M. R. Phase recovery and lensless imaging by iterative methods in optical, X-ray and electron diffraction. *Philos. Trans. R. Soc.* London, Ser. A **360**, 875–895 (2002).

22. Thibault P. & Rankenburg I. Optical Diffraction Microscopy in a Teaching Laboratory. *Am. J. Phys.*, **75** (9) 827-832 (2007).

23. Sandberg, R. L. *et al.* Lensless Diffractive Imaging Using Tabletop Coherent High-Harmonic Soft-X-Ray Beams. *Phys. Rev. Lett.* **99**, 098103 (2007).

24. Sandberg, R. L. *et al.* High Numerical Aperture Tabletop Soft X-ray Diffraction Microscopy with 70 nm Resolution. *Proc. Natl. Acad. Sci. USA* **105**, 24-27

(2008).

25. Chapman, H. N. *et al.* Femtosecond diffractive imaging with a soft-X-ray free-electron laser. *Nature Phys.* **2**, 839-843 (2006).

26. Barty, A. *et al.* Ultrafast single-shot diffraction imaging of nanoscale dynamics *Nature Photon.* **2**, 415-419 (2008).

27. Mancuso, A. P. *et al.* Coherent-Pulse 2D Crystallography Using a Free-Electron Laser X-Ray Source. *Phys. Rev. Lett.* **102**, 035502 (2009).

28. Miao J., Sayre D. & Chapman H. N. Phase retrieval from the magnitude of the Fourier transforms of non-periodic objects. *J. Opt. Soc. Am. A* **15**:1662-1669 (1998).

29. Bubeck, D. *et al.* The Structure of the Poliovirus 135S Cell Entry Intermediate at 10-Angstrom Resolution Reveals the Location of an Externalized Polypeptide That Binds to Membranes. *J. Virol*. **79**, 7745-7755 (2005).

30. Neutze, R., Wouts, R., Spoel D., Weckert, E. & Hajdu, J. Potential for biomolecular imaging with femtosecond X-ray pulses. *Nature* **406**, 752-757 (2000).

**Supplementary Information** is linked to the online version of the paper at www.nature.com/nature.

**Acknowledgements** We thank M. M. Murnane for stimulating discussions, C. Song for performing data analysis, Y. Mao for implementing an interpolation code, A. E. Sakdinawat for fabricating a test sample, P. Wachulak, M. Marconi, C. Menoni , J. J. Rocca and M. M. Murnane for help with data acquisition and T. Singh for parallelization of our phase retrieval codes. This work was supported by the U.S. DOE, Office of Basic Energy Sciences and the U.S. NSF, Division of Materials Research and Engineering Research Center, HHMI Gilliam fellowship for advanced studies and the UCLA MBI Whitcome fellowship. We used facilities supported by the NSF Center in EUV Science and Technology.

**Figure legends**

**Figure 1**. Schematic layout of the experimental set-up for ankylography. A coherent beam illuminates a finite object and the scattered waves form a diffraction pattern on the Ewald sphere (**a**), where $2\theta_{max}$ represents the diffraction angle. When the diffraction pattern is sampled at a scale sufficiently finer than the Nyquist interval, the 3D structure of the object is encoded into the 2D spherical pattern and can be directly reconstructed (**b**). Here, the 3D object is a sodium silicate glass particle with a size of 14×14×14 Å. The red, purple and yellow spheres represent the positions of O, Na and Si atoms, respectively.

**Figure 2**. 3D structure determination of a sodium silicate glass particle at 2 Å resolution from a simulated 2D spherical diffraction pattern alone. Electron density distribution of a 2-Å-thick slice of the MD simulation (**a**) and the reconstructed structure (**b**) along the XY plane where the red, purple and yellow spheres represent the positions of O, Na and Si atoms, respectively. Electron density distribution of a 2-Å-thick slice of the MD simulation (**c**) and the reconstructed structure (**d**) along the XZ plane. The width and height of each slice are 14 Å.

**Figure 3**. 3D structure determination of an individual poliovirus from a single simulated X-FEL pulse. **a**, Iso-surface rendering of the reconstructed viral

capsid structure, showing a fivefold mesa (red dots) and a threefold propeller tips (blue dots). The canyon between the mesa and the propeller structure (labelled with an arrow) is the expected binding site to its receptor. **b**, A 1.65-nm-thick central slice of the reconstructed 3D virus structure across the fivefold mesa and the propeller structure. The arrow indicates the receptor binding site. Scale bar is 5 nm.

**Figure 4**. Demonstration of ankylography using experimental data obtained with a soft X-ray laser. **a**, Oversampled 2D diffraction pattern on the Ewald sphere. Iso-surface renderings of the 3D reconstructed image in the XY plane (**b**) and XZ plane (**c**), respectively, where the incident beam is along the Z-axis. **d**, Lineout indicating the width of the tilted sample's projection onto the Z-axis to be 405 nm. **e**, SEM image of the sample (scale bar = 1 μm). Insets show two structure defects (labelled with arrows) in the sample which are spatially resolved in the 3D reconstructed image (arrows in **b**).

**FIGURES**

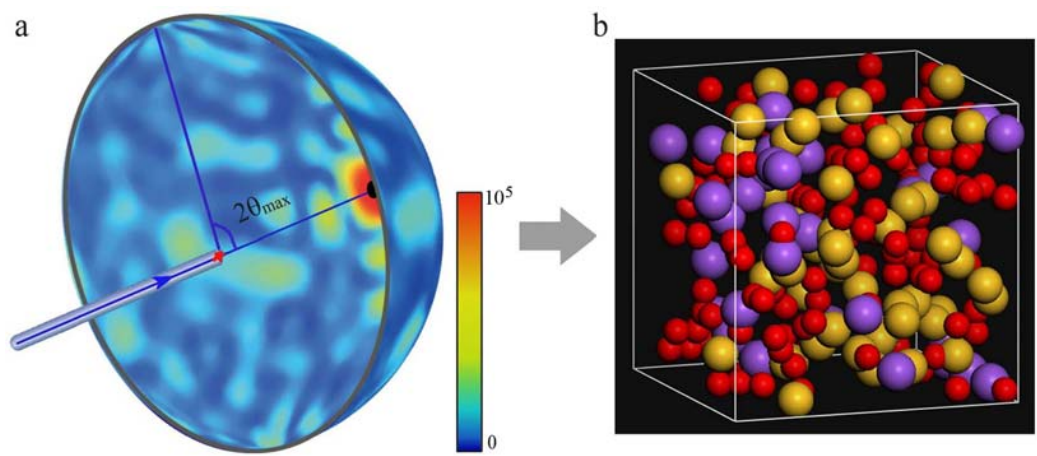

FIG. 1

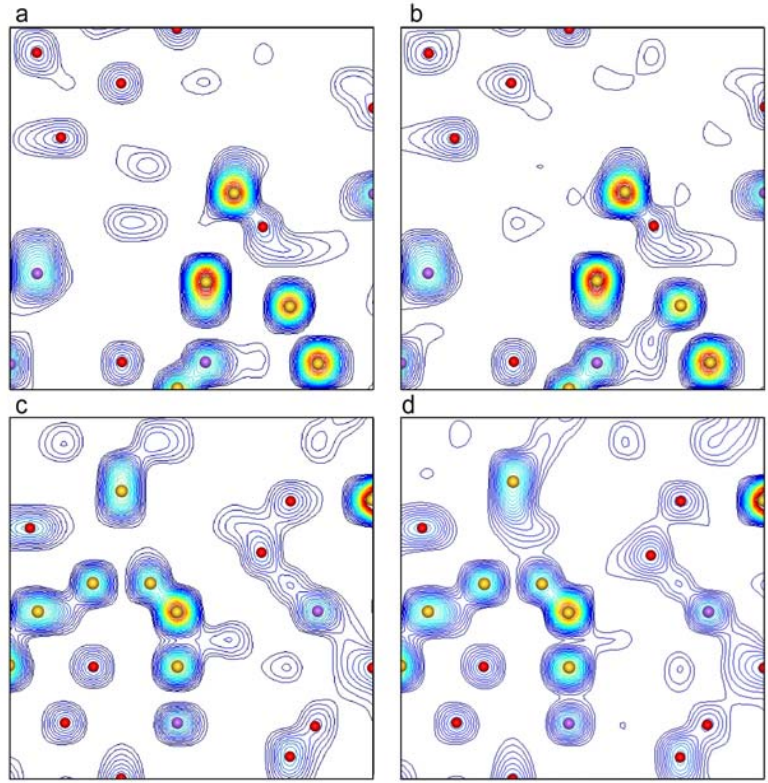

FIG. 2

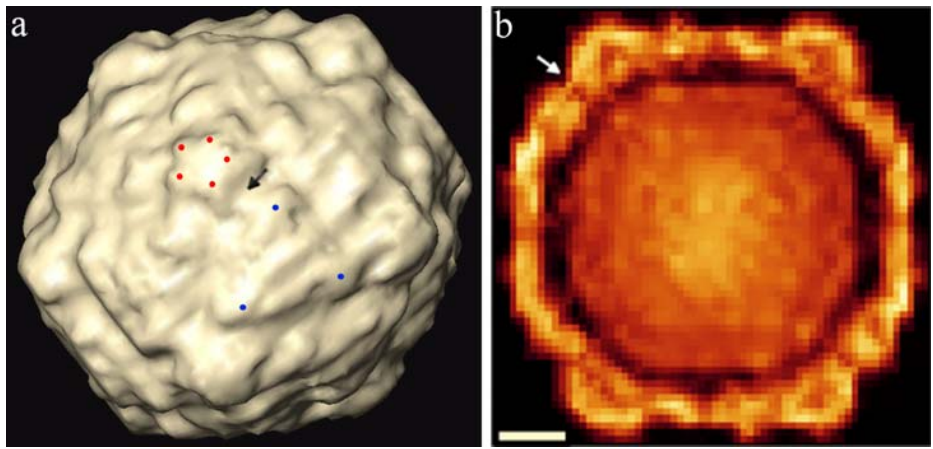

FIG. 3

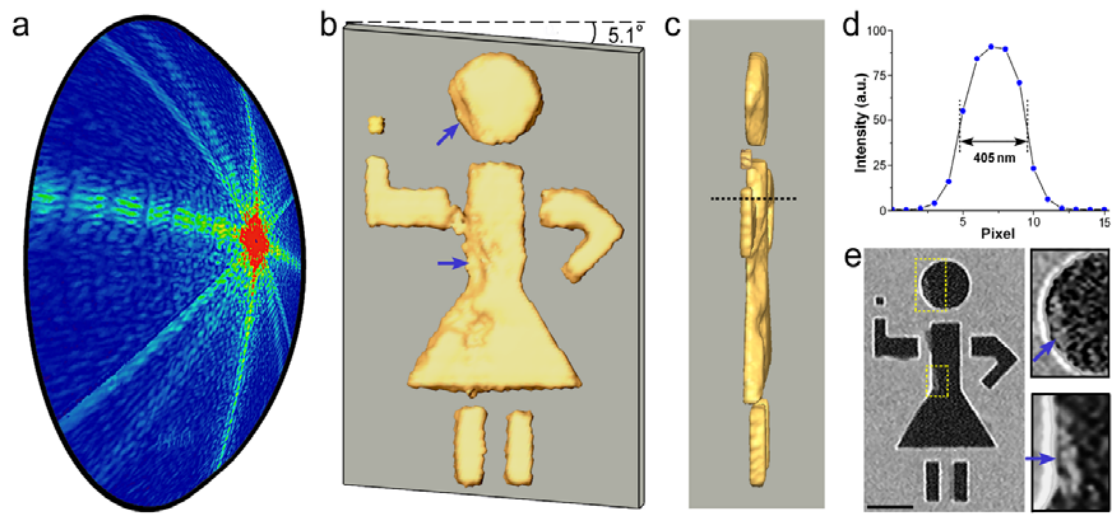

FIG. 4

# SUPPLEMENTARY INFORMATION

Kevin S. Raines[1], Sara Salha[1], Richard L. Sandberg[2], Huaidong Jiang[1], Jose A. Rodríguez[3], Benjamin P. Fahimian[1], Henry C. Kapteyn[2], Jincheng Du[4] and Jianwei Miao[1]

[1]*Department of Physics and Astronomy and California NanoSystems Institute, University of California, Los Angeles, CA 90095, USA.* [2]*Department of Physics and JILA, University of Colorado and NIST, Boulder, CO 80309, USA.* [3]*Molecular Biology Institute, University of California at Los Angeles, Los Angeles, CA 90095, USA.* [4]*Deparment of Materials Science & Engineering and Center for Advanced Scientific Computing and Modeling, University of North Texas, Denton, TX 76203, USA*

This file contains Supplementary Methods, Supplementary Figures, Supplementary Discussion and Supplementary References.

## Supplementary Methods

The $25Na_2O$-$75SiO_2$ glass structure was obtained by combining classical molecular dynamics (MD) simulations and subsequent geometry optimization using *ab initio* Density Functional Theory (DFT) calculations. Final electron density for both core and valence electrons was calculated using the DFT based on the optimized geometry. The simulation cell contains 204 atoms in a cubic box of 14×14×14 Å dimension to give the glass density of ~ 2.43 g/cm$^3$. The simulated melt and quench procedure was used to generate the glass structure by MD simulations with partial charge potentials as described elsewhere[31]. Starting from randomly generated initial atom positions, the system was first melted at 6000 K and then equilibrated at 4000 K, each with 100 ps NVT run and subsequent 100 NVE run. From 4000 K, the melt was then cooled down to 300 K by gradually decreasing the temperature at a cooling rate of 10 K/ps. At 300 K, the glass was equilibrated for 200 ps NVT runs and 200 ps NVE runs. Glass structure from MD simulation was further optimized using plane wave basis set, gradient corrected DFT calculations carried out using the Vienna *ab initio* Simulation Package (VASP) at highly parallel super computers. Projected Augmented Wave (PAW) pseudopoential and the PBE exchange and correlation function was used in the DFT calculations. Gamma point only K point sampling was used due to large simulation cell in our calculations. A plane wave kinetic energy cutoff of 400 eV was used. The convergence criterion for electronic SCF was 1e$^{-4}$ eV and geometry optimization was stopped when the forces acting on atoms were smaller than 0.01 eV/Å. Based on the optimized structure, total electron density (both core and valence electrons) were generated on a grid for subsequent image reconstruction.

**Supplementary Figures**

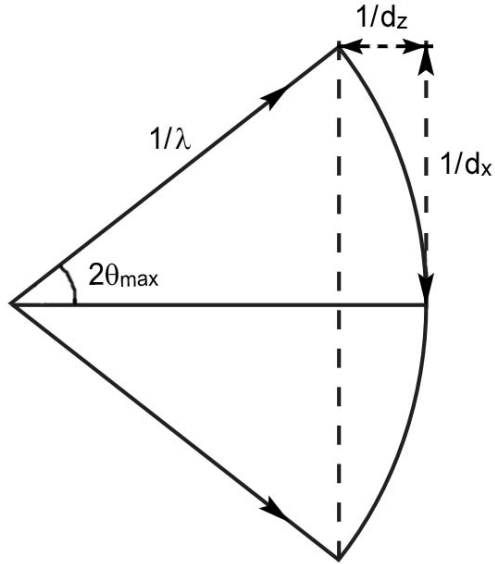

**Supplementary Fig. S1** Cross-section of a 2D diffraction pattern on the Ewald sphere, illustrating the 3D spatial resolution in ankylography. The resolution along the X and Z-axis are $d_x$ and $d_z$, respectively, $2\theta_{max}$ is the diffraction angle and $1/\lambda$ is the radius of the Ewald sphere. Eq. (2) in the manuscript was derived based on this geometrical construction and Bragg's law.

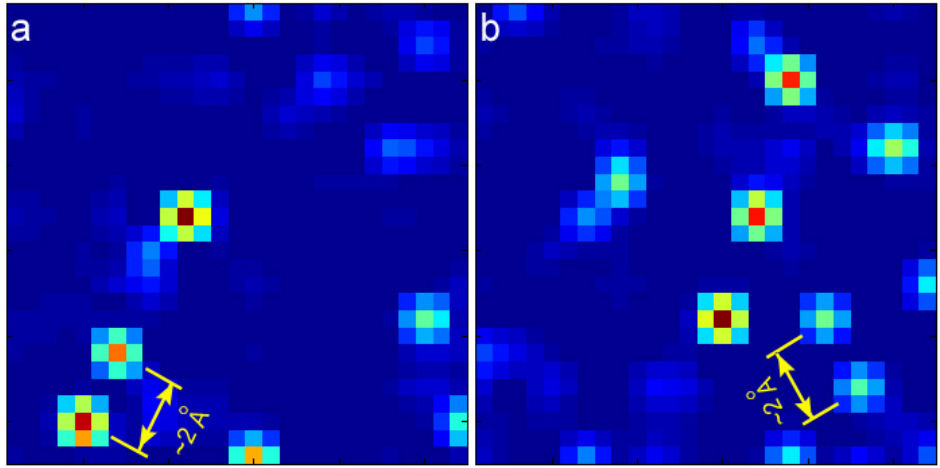

**Supplementary Fig. S2** Quantification of the 3D spatial resolution in ankylography with numerical simulations. **(a), (b)** Two 1-Å-thick slices of the reconstructed sodium silicate glass structure along the XY and XZ planes, respectively. By measuring the distance of nearest neighboring atoms, we confirmed the resolution along the XY and XZ planes to be 2 Å, which is consistent with the results calculated from Eq. (2).

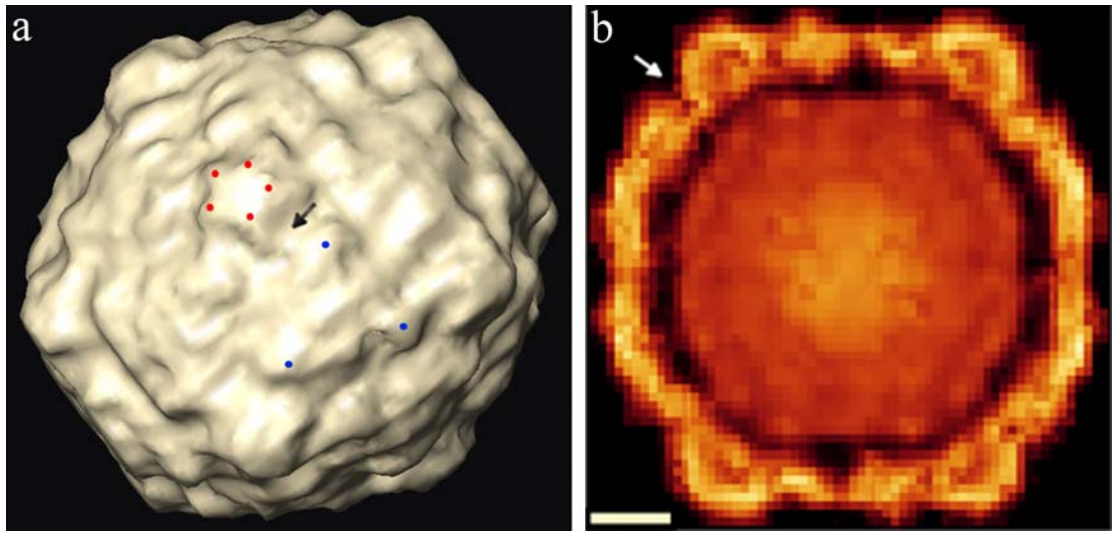

**Supplementary Fig. S3** 3D model structure of the poliovirus obtained by cryo-electron microscopy through averaging 8224 particles[32]. **a**, Iso-surface rendering of the viral capsid structure, showing a fivefold mesa (red dots) and a threefold propeller tips (blue dots). The canyon between the mesa and the propeller structure (arrow) is the expected binding site to its receptor[33]. **b**, A central slice of the model 3D virus structure across the fivefold mesa and the propeller structure. The arrow indicates the receptor binding site. Scale bar is 5 nm.

## Supplementary Discussion

**I. Representation of a 2D diffraction pattern on the Ewald sphere**

Based on the Born approximation, the 3D diffraction intensities of a finite object in the far field can be represented as

$$\left|F(\vec{k})\right| = \left|\int_V \rho(\vec{r})e^{-2\pi i \vec{k}\cdot\vec{r}} d^3\vec{r}\right| \qquad (S1)$$

where $\left|F(\vec{k})\right|$ is proportional to the square root of the diffraction intensities, $\vec{k}=(k_x,k_y,k_z)$ is the reciprocal-space vector, $\rho(\vec{r})$ is the 3D structure of the object with $\vec{r}=(x,y,z)$, and $V$ is the spatial extent of the object. To calculate the diffraction intensities on the Ewald sphere, we constructed a geometrical representation of the Ewald sphere, shown in Fig. S4. By expressing $\vec{k}$ in the spherical polar coordinates, we have

$$k_x = S_x = \frac{1}{\lambda}\sin\theta\cos\phi$$

$$k_y = S_y = \frac{1}{\lambda}\sin\theta\sin\phi \qquad (S2)$$

$$k_z = S_z - S_0 = \frac{1}{\lambda}(\cos\theta - 1)$$

where $\vec{S}_0$ and $\vec{S}$ are the incident and scattered wave vectors with $\vec{S}=(S_x,S_y,S_z)$, and $\theta$

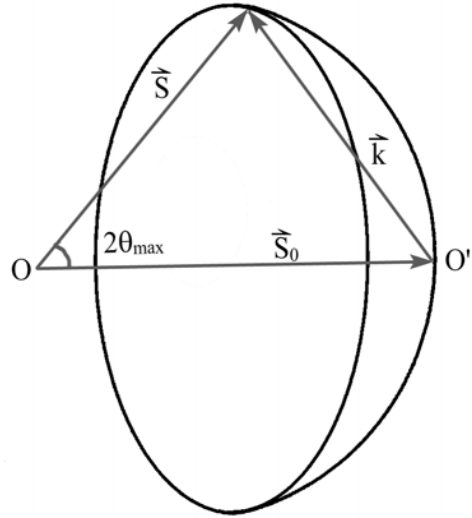

**Supplementary Fig. S4** Geometrical construction of the wave vectors and the Ewald sphere. $\vec{S}_0$ and $\vec{S}$ are the incident and scattered wave vectors with $\left|\vec{S}_0\right|=\left|\vec{S}\right|=\frac{1}{\lambda}$, and $\lambda$ is the wavelength of the incident beam. O is the origin of the Ewald sphere and O' is the origin of $\vec{k}$.

and $\varphi$ are the spherical polar coordinates. By substituting Eq. (S2) into Eq. (S1), we obtained Eq. (1) in the manuscript.

## II. An oversampled 2D diffraction pattern on the Ewald sphere determines the 3D structure of a finite object

Here we present a qualitative explanation of why an oversampled 2D diffraction pattern on the Ewald sphere determines the 3D structure of a finite object. For the purpose of clarity, we focus on a real-valued object, but our arguments remain valid for the complex case as well. Fig. S5 shows an oversampled diffraction pattern on the Ewald sphere with the diffraction angle $2\theta_{max} = 90°$. The presence of the diffraction patterns on two hemispheres comes from the centro-symmetry of the intensities. The question of uniqueness can be addressed through the relationship of the Ewald sphere to a central projection. According to the Fourier slice theorem, the Fourier transform of a projection of a 3D object at a given angle is equal to a slice through the origin of the 3D Fourier transform of the object[34]. Fig. S5 shows a central Fourier slice intersecting with the diffraction pattern on the Ewald sphere and forming a circle. When the 3D object is rotated around the X-axis, the intersection between the Ewald sphere and the corresponding central Fourier slice will either form a circle, an arc or a point, which can be expressed as

$$e_\alpha = D_E \cap P_\alpha \qquad (S3)$$

where $e_\alpha$ represents the intersection at a given rotation angle $\alpha$ along the X-axis, $\cap$ is the intersect operation, $D_E$ the diffraction pattern on the Ewald sphere and $P_\alpha$ a central Fourier slice at angle $\alpha$. When the diffraction pattern is sufficiently oversampled with the oversampling degree $O_d > 1$, $D_E$ is the set consisting of the intersected arcs, circles and point along the all possible rotation angles about the X-axis,

$$D_E = \{0 \le \alpha < 180°: \quad D_E \cap P_\alpha\} \qquad (S4)$$

We have so far restricted the discussion to the case of rotation about the X-axis.

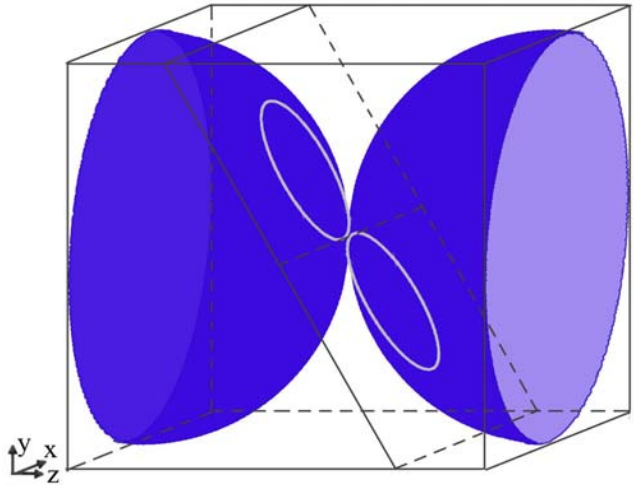

**Supplementary Fig. S5** Intersection between a Fourier slice and the Ewald sphere at a give angle, forming a circle, an arc or a point. The two hemispheres are due to the centro-symmetry of the diffraction intensities and the X-axis is the rotation axis.

However, we note that the oversampled diffraction pattern on the Ewald sphere can be constructed by rotating about either X, Y or Z-axis, where the argument is repeated. We hence have

$$D_E = \left\{ 0 \leq \alpha, \beta, \gamma < 180°: \quad D_E \cap P_\alpha, D_E \cap P_\beta, D_E \cap P_\gamma \right\} \quad \quad (S5)$$

where $\beta$ and $\gamma$ are the rotation angles about the Y and Z-axis, respectively. Eq. (S5) indicates that the oversampled diffraction pattern on the Ewald sphere is the set consisting of the intersected arcs, circles and point in *all possible* orientations. If two 3D objects share exactly the same oversampled diffraction pattern on the Ewald sphere, they must possess the identical intersected arcs and circles of the central Fourier slices in all possible orientations. Thus the two structures must be equivalently correlated for all possible projections and any difference in the structures that may be concealed at one orientation must be equivalently concealed for all orientations. On this basis of equivalent projections and our numerical results, we determine then that the Ewald sphere constraint is sufficiently strong as to generally determine the complete 3D structure.

We conclude that an oversampled diffraction pattern on the Ewald sphere, in combination with real-space constraints, suffices to uniquely determine general 3D structures for all system sizes of practical interest. Although we don't provide a rigorous mathematical proof, we have performed a variety of numerical simulations showing that the 3D object is not only unique but also recoverable – a more difficult condition to fulfill.

**III. Spatial and temporal coherence required in ankylography**

The required spatial and temporal coherence in ankylography are determined by

$$\Delta\theta \leq \frac{\lambda}{2\pi\sigma_i a} \quad \quad \text{and} \quad \quad \frac{\lambda}{\Delta\lambda} \geq \frac{\sigma_i a}{d} \quad \quad (S6)$$

where $\Delta\theta$ the divergence or convergence angle of the incident beam, $\sigma_i$ the linear oversampling ratio defined in ref. 35, $a$ the sample size, $\Delta\lambda$ the wavelength spread of the beam, and $d$ the desired resolution.

**IV. The iterative algorithm**

We introduce three new constraints in the ankylography reconstructions to accommodate the large number of missing data and the large no-density region: uniformity outside of the support region, continuity inside the support, and amplitude extension in Fourier space.

Uniformity outside the support

Since the 2D spherical diffraction pattern in ankylography needs to be finely oversampled, a large accompanying region outside of the support is necessarily present in the reconstructions. In principle, this region should be zero. However, at the early stages of convergence, and particularly when noise is present in the measured diffraction intensities, this constraint is generally not satisfied. In the hybrid-input-output (HIO) algorithm[36], one takes advantage of this knowledge by iteratively suppressing all non-zero values outside the support; here we supplement this constraint by also exploiting the fact that the region outside the support should be *uniform*. That is, since we know that all image estimates will not entirely satisfy the support constraint, we not only push the non-zero values towards zero, but also impose uniformity upon them as well. This constraint is enforced by convolving the density outside the support with a unit step function over a 3×3×3 voxel box. The purpose of this constraint is to reduce the fluctuations outside the support and make the convergence smoother.

Continuity inside the support

One common problem in phase retrieval is the stagnation within local minima, hence we introduce a perturbation into the reconstruction, which is designed to bump the reconstruction out of any minimum in which it may be confined in order for the algorithm to continue its search for a global minimum. We achieve this by periodically smoothing the recovered density with a Gaussian filter. The reason for a Gaussian filter inside the support as opposed to the box function outside the support is due to the fact that the Gaussian filter only imposes a continuity constraint upon the data, whereas the box filter enforces uniformity. In our reconstructions, we used a normalized Gaussian filter with a standard deviation of 0.65 sampled within a 3x3x3 array.

Amplitude extension

In order to capitalize on the fast convergence of small array reconstructions, where knowledge of a loose support (*i.e.* a boundary somewhat larger than the true sample envelope) is sufficient to reconstruct the image, we first compute a reconstruction from the low-resolution diffraction pattern on the Ewald sphere. We then compute a higher resolution reconstruction, including both the diffraction pattern on the Ewald sphere and the recovered Fourier amplitudes at low resolution. We repeat this procedure until the reconstruction of the full 3D array is reached. In each amplitude extension, we usually increase the number of unknown density elements in real space by a factor of between 2 and 5.

Although we have only implemented the amplitude extension constraint here, we note that there are a few more constraints that can be computed at low resolution to assist the high resolution reconstruction. These constraints include but are not limited to: phase extension[37], histogram matching[38], and molecular replacement[39]. Our preliminary simulations to date indicate that enforcing these constraints for large system sizes dramatically improves the reconstruction convergence and will be the subject of subsequent publications.

The complete iterative algorithm

The final description of our algorithm includes how often and in what order these constraints are implemented. Below we list the detailed steps involved in the reconstruction process.

1. Start with 50 independent reconstructions using random initial phase sets and an updated support (for the first iteration, a loose support is used which is estimated from the speckle size).
2. Determine the best random initial phase set based on the error metric in the Fourier space at the end of 1000 HIO iterations. β in the HIO algorithm is set to be 0.9 (ref. 36).
3. Use the best random initial phase set and compute 1000 iterations of HIO.
4. Continue HIO by periodically applying the uniformity constraint once every 10 iterations to the region outside the support until the error stagnates.
5. Apply the continuity constraint to the region inside the support.
6. Repeat steps 3-5 until the error stagnates.
7. Compute a tight support by smoothing and thresholding the reconstruction.
8. Repeat steps 1 – 7 until the best reconstruction is obtained.

## V. 3D ankylographic reconstructions

By using the iterative algorithm described above, we have computed 3D reconstructions of a sodium silicate glass structure, a poliovirus and a test sample from the experimental data. Among the three reconstructions, the poliovirus is the most complicated. To better illustrate our reconstruction procedures, we first describe the virus reconstruction, followed by the glass structure and the test sample reconstructions.

3D reconstruction of a single poliovirus

We obtained the poliovirus structure from the Electron Microscopy Data Bank[32] (http://www.ebi.ac.uk/pdbe/emdb/ entry 1136). To obtain a noisy 2D spherical diffraction pattern, we simulated an X-FEL pulse with λ = 1.77 nm and $10^{13}$ photons per pulse that was focused down to a spot of 100 nm, and illuminated a single poliovirus particle. A diffraction pattern on the Ewald sphere was calculated with $2\theta_m$ = 62.6°, which corresponds to $q_{max}$ = 3.7 nm$^{-1}$. Poisson noise was added to the diffraction intensities. To simulate a beam stop, we removed the intensities at the central 7×7 pixels. Fig. S6 shows a front view of the spherical diffraction pattern with a dynamic range between 0 and $1.4\times10^6$ photons.

We computed the reconstruction of the poliovirus by using the iterative algorithm described above. Table S1 lists the parameters used in the algorithm. We began with a small array corresponding to a maximum diffraction angle of 16.1°, and then subsequently computed four additional reconstructions through amplitude extension. In Tab. S1, we characterize each step in the reconstruction process by a series of measures:

*R-space Array*: The size of the smallest box that completely encloses the virus density.
*K-space Array*: The size of the array containing the embedded Ewald sphere.
$O_d$ (oversampling degree): The ratio of the number of points on the Ewald sphere to the number of voxels within a support.
$O_t$ (total oversampling degree): The ratio of the total number of points in Fourier space (points on the Ewald sphere and points calculated through the amplitude extension method) to the number of voxels within the support.

*Diffraction Angle*: The maximum diffraction angle of the current reconstruction.

$R_{Ewald}$: The R-factor of the reconstruction along the Ewald sphere, defined as

$$R_{Ewald} = \sum \left| |F^T_{Ewald}(\vec{k})| - |F^R_{Ewald}(\vec{k})| \right| \Big/ \sum \left| F^T_{Ewald}(\vec{k}) \right|, \text{ where } |F^T_{Ewald}(\vec{k})| \text{ and } |F^R_{Ewald}(\vec{k})|$$ are the true and reconstructed Fourier magnitudes on the Ewald sphere.

$R_{Entire}$: The R-factor of the reconstruction over the entire K-space.

| R-space Array | K-space Array | $O_t$ | Diffraction Angle | $R_{Ewald}$ | $R_{Entire}$ |
|---|---|---|---|---|---|
| 10×10×2 | 40×40×6 | 7.29 | 16.1° | 0.0004 | 0.002 |
| 16×16×4 | 64×64×16 | 7.34 | 26.3° | 0.0007 | 0.010 |
| 20×20×6 | 80×80×24 | 14.58 | 33.7° | 0.002 | 0.015 |
| 24×24×10 | 96×96×38 | 13.99 | 41.7° | 0.002 | 0.021 |
| 32×32×20 | 128×128×78 | 9.07 | 62.6° | 0.003 | 0.14 |

Tab. S1 Parameters and R-factors used in the iterative algorithm for the reconstruction of the poliovirus structure.

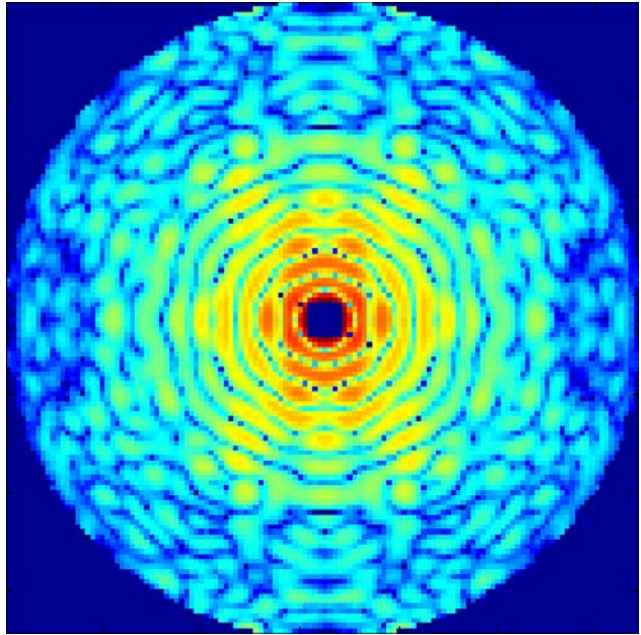

**Supplementary Fig. S6** Front view of a noisy diffraction pattern on the Ewald sphere, obtained from a single poliovirus. The intensities at the central 7×7 pixels were removed. The dynamic range of the intensities is from 0 to $1.4 \times 10^6$ photons.

Due to the discrete nature of the model poliovirus structure, it is impossible to exactly simulate the continuous diffraction intensities upon the Ewald sphere (*i.e.* errors in discrete Fourier transform). To resolve this issue, we first interpolated the poliovirus structure into a 10×10×2 voxel array (Tab. S1), and calculated its oversampled Fourier magnitudes on the Ewald sphere with Poisson noise (a 40×40×6 voxel array, denoted $A_1$). By using the iterative algorithm, we reconstructed the low resolution virus structure as well as the entire Fourier magnitudes, $|F_{Recon}^{A_1}(\vec{k})|$. We then interpolated the poliovirus structure into a 16×16×4 voxel array, and calculated its Fourier magnitudes on the Ewald sphere with Poisson noise, $|F_{Ewald}^{A_2}(\vec{k})|$, where $A_2$ represents a 64×64×16 voxel array. To implement amplitude extension, we calculated the hybrid Fourier magnitudes by

$$|F_{Hybrid}^{A_2}(\vec{k})| = \begin{cases} \omega \cdot |F_{True}^{A_2}(\vec{k})| + (1-\omega) \cdot \dfrac{|F_{Recon}^{A_1}(\vec{k})|}{|F_{True}^{A_1}(\vec{k})|} \cdot |F_{True}^{A_2}(\vec{k})| & \vec{k} \in A_1 \\ |F_{Ewald}^{A_2}(\vec{k})| & \vec{k} \in A_2 \end{cases} \quad (S7)$$

where $|F_{True}^{A_1}(\vec{k})|$ and $|F_{True}^{A_2}(\vec{k})|$ represent the entire Fourier magnitudes calculated from the 10×10×2 voxel and 16×16×4 voxel poliovirus structures, respectively, and $\omega$ is a parameter between 0 and 1. We adjusted $\omega$ such that

$$\frac{\sum \left||F_{Hybrid}^{A_2}(\vec{k})| - |F_{True}^{A_2}(\vec{k})|\right|^2}{\sum |F_{True}^{A_2}(\vec{k})|^2} = \frac{\sum \left||F_{Recon}^{A_1}(\vec{k})| - |F_{True}^{A_1}(\vec{k})|\right|^2}{\sum |F_{True}^{A_1}(\vec{k})|^2} \quad \vec{k} \in A_1 \quad (S8)$$

The assembled $|F_{Hybrid}^{A_2}(\vec{k})|$ was then used to reconstruct the 16×16×4 voxel poliovirus structure. We repeated the amplitude extension procedure (Tab. S1) until the full-size reconstruction of the poliovirus structure was completed. Note that for experimental data, it is unnecessary to use Eqs. (7) and (8).

3D reconstruction of a sodium silicate glass structure and a test sample

For smaller size arrays or thin objects, such as the glass structure and the test sample described in the manuscript, one may relax the amplitude extension and still recover the object density. For the sodium silicate glass structure described above, a noisy diffraction pattern was obtained from a simulated Energy Recovery Linac beam with λ = 2 Å and a coherent flux of $10^{14}$ photons/s that was focused down to a beam spot size of 100 nm and illuminated the sodium silicate glass particle[40]. A diffraction pattern on the Ewald sphere was recorded with $2\theta_m$ = 90° and an exposure time of 5.1 s, which corresponds to $q_{max}$ = 4.4 Å$^{-1}$. Poisson noise was added to the diffraction intensities, shown in Fig. 1a.

We computed the reconstruction of the sodium silicate glass structure using the iterative algorithm described above. Since the array size in real space is relatively small, we didn't use the amplitude extension method. The other steps are similar to those used for the poliovirus reconstructions. Table. S2 shows the parameters for the reconstruction process. The small R-factors indicate the robustness of the reconstruction. Compared to the original structure, the atomic positions are clearly resolved in the reconstruction (Figs. 2a-d).

Similarly, we followed the same procedures in reconstructing the experimental data of the test sample (Tab. S2). To verify the results, we conducted a few independent reconstructions and the reconstructed 3D images are very consistent. Note that we designated the spherical diffraction pattern by two angles. Due to the projection of a planar detector onto the Ewald sphere (see next section), the resolution along the Z-axis was calculated from the maximum angle extended at the corner of the pattern (48.3°); whereas the resolution along the X and Y axes is specified by the angle formed at midplane (35.9°); which yielded 80 nm and 140 nm resolutions, along the XY plane and the Z axis, respectively.

|  | R-space Array | K-space Array | $O_d$ | Diffraction Angle | $R_{Ewald}$ | $R_{Entire}$ |
|---|---|---|---|---|---|---|
| Glass structure | 14×14×14 | 64×64×64 | 2.7 | 90° | 0.005 | 0.112 |
| Test sample | 100×170×7 | 420×420×240 | 2.6 | 35.9° | 0.08 | NA |

Tab. S2 Parameters and R-factors used in the iterative algorithm for the reconstructions of the sodium silicate glass structure and the test sample

## VI. Interpolation of a planar diffraction pattern onto the Ewald sphere

Since the diffraction pattern was measured by a planar CCD detector, in order to perform 3D phase retrieval, the diffraction intensities have to be normalized[41] and then projected onto the Ewald sphere. As Fig. S7 shows, each pixel of the planar detector integrates the diffracted photons within a solid angle, $\Delta\Omega(\vec{k})$, subtended by the pixel area. Since $\Delta\Omega(\vec{k})$ decreases with the increase of the spatial frequency, we normalize the diffraction intensities by

$$I(\vec{k}) = \frac{I_M(\vec{k})}{\Delta\Omega(\vec{k})} \qquad (S9)$$

where $I(\vec{k})$ and $I_M(\vec{k})$ are the normalized and measured intensities.

After the normalization, the diffraction intensities have to be projected onto the Ewald sphere. Based on the geometry shown in Fig. S7, we obtain the relationship between the vector on the Ewald sphere and the vector on the detector plane,

$$k_x = \frac{Rk'_x}{\sqrt{R^2 + k'^2_x + k'^2_y}}$$

$$k_y = \frac{Rk'_y}{\sqrt{R^2 + k'^2_x + k'^2_y}} \quad (S10)$$

$$k_z = R\left(1 - \frac{R}{\sqrt{R^2 + k'^2_x + k'^2_y}}\right)$$

where $\vec{k} = (k_x, k_y, k_z)$ is a vector on the Ewald sphere, $\vec{k}' = (k'_x, k'_y)$ is a vector on the 2D planar detector, and $R$ the distance from the sample to the detector. As Fig. S7 shows, $k'_x$ and $k'_y$ are on a regular 2D grid, but $k_x$, $k_y$ and $k_z$ are on an irregular 3D grid. Since a regular 3D grid in Fourier space is mandatory for the phase retrieval algorithm due to the use of the FFT, we interpolate $I(\vec{k})$ onto a regular 3D Cartesian grid by using a linear interpolation method[34].

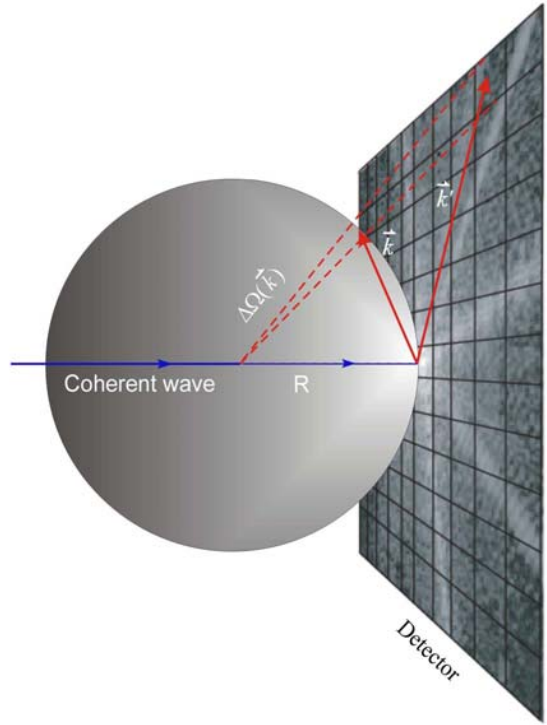

**Supplementary Fig. S7** Projection of an oversampled 2D diffraction pattern from a planar detector onto the Ewald sphere, where $\vec{k} = (k_x, k_y, k_z)$ is a vector on the Ewald sphere, $\vec{k}' = (k'_x, k'_y)$ a vector on the detector plane and $\Delta\Omega(\vec{k})$ a solid angle subtended by the pixel area.

## Supplementary References

31. Du, D. & Corrales, L. R. Compositional dependence of the first sharp diffraction peaks in alkali silicate glasses: A molecular dynamics study. *J. Non-Cryst. Solids*, **352**, 3255-3269 (2006).
32. Bubeck, D. *et al.* The Structure of the Poliovirus 135S Cell Entry Intermediate at 10-Angstrom Resolution Reveals the Location of an Externalized Polypeptide That Binds to Membranes. *J. Virol*. **79**, 7745-7755 (2005).
33. Belnap, D. M. *et al*. Three-dimensional structure of poliovirus receptor bound to poliovirus. *Proc. Natl. Acad. Sci. USA* **97**, 73-78 (2000).
34. Kak, A.C. & Slaney, M. *Principles of Computerized Tomographic Imaging.* (SIAM, Philadelphia, 2001).
35. Miao, J. *et al*., Quantitative Image Reconstruction of GaN Quantum Dots from Oversampled Diffraction Intensities Alone. *Phys. Rev. Lett*. **95**, 085503 (2005).
36. Fienup, J. R. Reconstruction of an object from the modulus of its Fourier transform. *Opt. Lett.* **3**:27-29 (1978).
37. Vellieux, F. M. D. & Read, R. J. Noncrystallographic symmetry averaging in phase refinement and extension, *Methods in Enzymol.* **277**, 18-53 (1997).

38. Zhang, K. Y. J. & Main, P. The Use of Sayre's Equation with Solvent Flattening and Histogram Matching for Phase Extension and Refinement of Protein Structures. *Acta Cryst*. A**46**, 377-381 (1990).
39. Argos, P & Rossman, M. G. Molecular replacement method. *In Theory and Practice of Direct Methods in Crystallography*, ed. MFC Ladd, RA Palmer, pp. 361-417. New York: Plenum (1980).
40. Bilderback, D. H., Elleaume, P. & and Weckert, E. Review of Third and Next Generation Synchrotron Light Sources. *J. Phys. B: At. Mol. Opt. Phys*. **38**, S773-S797 (2005).
41. Song, C. *et al*. Phase Retrieval from Exactly Oversampled Diffraction Intensity Through Deconvolution. *Phys. Rev. B*. **75**, 012102 (2007).